\documentclass[10pt,a4paper]{article}
\usepackage{amsmath,amssymb}
\usepackage{url}
\usepackage{placeins}
\usepackage{subfigure}
\usepackage{multirow}
\usepackage{epsfig}
\usepackage{chicago}

\newcommand{\h}{\ensuremath{\frac{1}{2}}}

\newcommand\be{$$}
\newcommand\ee{$$}
\newcommand\ben{\begin{equation}}
\newcommand\een{\end{equation}}
\newcommand\bea{\begin{eqnarray*}}
\newcommand\eea{\end{eqnarray*}}
\newcommand\bean{\begin{align}}
\newcommand\eean{\end{align}}

\author{Chris Kenyon and Andrew Green\footnote{Contact: chris.kenyon@lloydsbanking.com}}
\title{Collateral-Enhanced Default Risk\footnote{\bf The views expressed are those of the authors only, no other representation should be attributed.}
}
\date{\today\\ \vskip5mm Draft 1.00}

\begin{document}

\maketitle

\begin{abstract}
Changes in collateralization have been implicated in significant default (or near-default) events during the financial crisis, most notably with AIG.  We have developed a framework for quantifying this effect based on moving between Merton-type and Black-Cox-type structural default models.  Our framework leads to a single equation that emcompasses the range of possibilities, including collateralization remargining frequency (i.e. discrete observations).  We show that increases in collateralization, by exposing entities to daily mark-to-market volatility, enhance default probability.  This quantifies the well-known problem with collateral triggers.  Furthermore our model can be used to quantify the degree to which central counterparties, whilst removing credit risk transmission, systematically increase default risk.  
\end{abstract}

\section{Introduction}

Currently there is a push from regulators and lawmakers to make banking safer and one consequence is to have as many trades as possible centrally cleared where they will be collateralized with standard terms \cite{BCBS-189,BCBS-206,BCBS-226,BCBS-227}.  Whilst uncollateralizd trades have always been penalized in terms of capital requirements, the required capital is increasing with Basel III \cite{BCBS-189}.  However, collateralization only mitigates credit risk transmission not default risk itself.   In fact we show that collateralization can significantly increase default risk because it exposes the participants to daily, and even intra-day, mark-to-market price volatility. We quantify this collateral-induced default using structural models. Thus central clearing and collateralization may make banking safer from transmission risk and less safe from default risk.

Changes in collateralization have been implicated in significant default or near-default events during the financial crisis, most notably in the case of AIG.  According to \cite{FCIC-aig-2010a} on September 15, 2008 AIG was downgraded and the total collateral calls increased from \$23.441B the day before to \$32.013B with AIG posting \$19.573B (an increase of less than \$1B from the previous day). On September 16 FRBNY annouced an \$85B loan to AIG which did not then default.  

Previous authors have pointed out that collateralization transforms transmission risk into liquidity risk \cite{Gregory2012a} because collateral must be obtained when required.  We show an accompanying transformation from transmission risk into default risk directly.  \cite{Kenyon2012a} described the collateral death spiral when assets decline in value requiring the holder to post more of them into collateral.  Having more assets tied up in collateral reduces the recovery rate of the institution which is then seen as a greater credit risk.  This leads depositors to remove their deposits further reducing the institution's recovery rate, and so on.  Another mechanism, this time related to the capital cost of not protecting trades has also been called the "doom loop" \cite{Murphy2012a} and is hypothesized to be one reason for the proposed exclusion of Sovereign exposures from at least one draft of CRD IV \cite{Cameron2012b}.  Our structural approach quantifies the results of these mechanisms rather than their dynamics directly.

The outline of the paper is as follows: we next describe collateralization then our modeling approach.  In the Results section we study some test cases and conclude in the Discussion section.

\section{Collateralization}

Bilateral (between two counterparties) collateral agreements are described in Credit Support Annexes (CSAs) to Master Agreements; ISDA provides standard templates.  CCP CSAs are produced by the CCP, e.g. LCH.  It is now recognized that CSAs themselves are highly complex derivatives \cite{Piterbarg2012a,ISDA-2011-CSA-overview}, hence the standardization initiative of ISDA with the Standardized-CSA (SCSA).  The SCSA also attempts to reflect CCP practices in pricing, e.g. specifying OIS discounting for derivative pricing.

The main features of a CSA are:
\begin{itemize}
	\item eligible products;
	\item eligible trade currencies: single defined currency or list;
	\item collateral currency: single or list;
	\item initial margin;
	\item remargining frequency;
	\item variation margin calculation;
	\item threshold amount;
	\item minimum transfer amount;
	\item independent amount;
	\item triggers and consequences (typically ratings-based);
	\item eligible collateral, haircuts on non-cash collateral;
	\item calculation agent (which party calculates the prices);
	\item dispute resolution mechanism.
\end{itemize}
The most notorious CSA feature is triggers which typically increase the collateral is response to the downgrade of one party.  This increase may change a situation from uncollateralized to fully collateralized (e.g. AIG).  
\begin{quote}
American International Group Inc. was facing a severe cash crunch last night as ratings agencies cut the firm's credit ratings, forcing the giant insurer to raise \$14.5 billion to cover its obligations…. 

AIG has been scrambling to raise as much as \$75 billion to weather the crisis, and people close to the situation said that if the insurer doesn't secure fresh funding by Wednesday, it may have no choice but to opt for a bankruptcy-court filing. 

(Wall Street Journal, 16 September 2008)
\end{quote}
Alternatively it may require posting an independent amount which is unrelated to specific trade details, say a flat 100M.

\section{Modeling}

We consider two key features of collateralization:
\begin{itemize}
	\item threshold
	\item frequency
\end{itemize}
By assumption we work in a single currency.  We include scenarios where there are downgrade triggers that depend on the volatility of firm value as a proxy for decreasing creditworthiness.  From a capital perspective all triggers must be ignored \cite{BCBS-189}.  This is because they will act procyclically, i.e. require a company to post more collateral when it is weaker.  However, triggers will still have material effects.  Some regulations require banks to hold and report liquidity buffers to cover effects of ratings downgrades (e.g. BIPRU 12.5 and related forms, FAS048, FAS050).  The key point is the timescale for the subsequent replenishment of those buffers.  We do not go into that level of detail.

\begin{figure}
	\centering
		\includegraphics[width=0.99\textwidth]{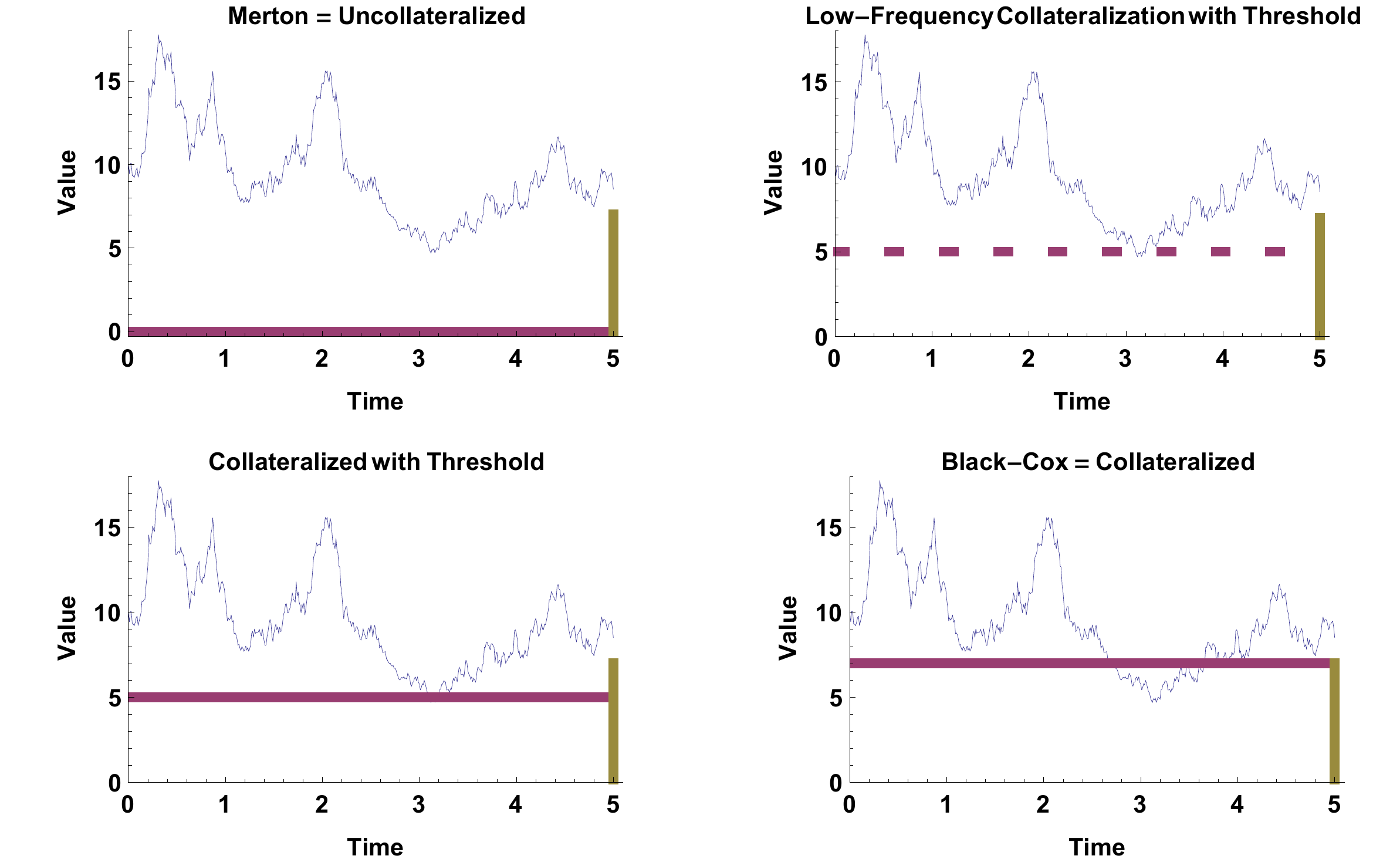}
	\caption{Example of default barriers from different models.  Default occurs if a barrier is touched.  For example in the uncollateralized (Merton) case (top left plot), there is no requirement to compare firm's assets with liabilities untill the last date (final vertical barrier).  In the fully collateralized (Black-Cox) case (bottom right) the firm's assets are compared continuously with liabilities.  Intermediate collateralization cases are given in the other plots (see text for more details).  The the horizontal barriers are lower than final (vertical) barrier in the first three plots to represent partial collateralization (e.g. with a threshold).  In the top two plots there would be no default whereas default would occur in the bottom two plots.}
	\label{f:examples}
\end{figure}

\begin{figure}
	\centering
		\includegraphics[width=0.5\textwidth]{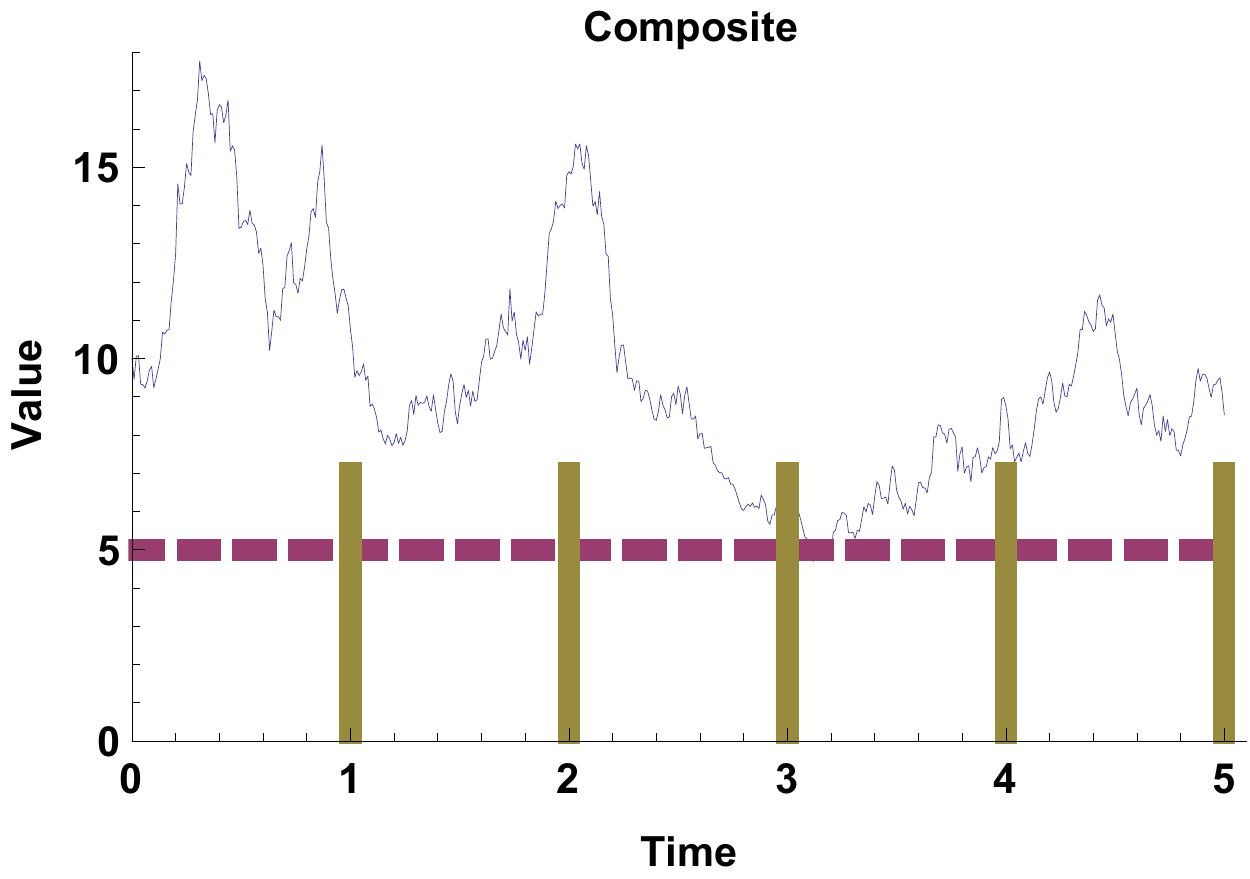}
	\caption{Composite barrier model where there are different barriers at different frequencies.  For example Accounting reports at quarter ends and weekly partial collateralization, thus mixing solvency with liquidity.  The Composite model can include any number of discrete barriers at different levels.}
	\label{f:composite}
\end{figure}

We base our analysis on a combination of two standard structural-default approaches, the Merton model \cite{Merton1974a} and the Black-Cox model \cite{Black1976a}.  These are described in terms of specific option positions below in the bullet points.  We update the Black-Cox model using \cite{Broadie1997a} to deal with discrete sampling, that is, collateral remargining at different frequencies.  Remargining is usually done daily, but some clearing houses (e.g. LCH) may also do intra-day remargining, whereas other entities may only remargin weekly or monthly.   Figure \ref{f:examples} shows an example path of firm asset value versus the default barriers in the different cases.

Our Composite model, shown in Figure \ref{f:composite} has different degrees of collateralization at different frequencies.  For example full collateralization at quarter ends and partial collateralization at the end of each week, or day.  The Composite model could be used, for example, for quantifying the default effect of weekly remargining and quarterly accounting.  Notice that this example would mix liquidity induced default (failure to post collateral because of insufficient assets) with solvency induced default (quarterly accounts showing assets below liabilities).

We can analyse survival in the structural default models above using standard (barrier) options as follows:
\begin{itemize}
	\item Merton: $P\{\tau>T\}$ = Binary European Call(K) / df(T)
	
	where df(T) is the discount factor to maturity (T)
	
	Survival in the Merton model occurs when the firm asset value is above a given level (the option strike, K) at a given time (option maturity, T).  To get the probability we have to remove the effect of the discount factor.
	
	\item Black-Cox: $P\{\tau>T\}$ = down-and-out(K) Binary Call(K) / df(T)
	
	Here the barrier level of the down-and-out barrier (K) is set to the same level as that of the Binary call.  Payment on the option only happens at maturity so we can recover the survival probability using the known discount factor to maturity as before.
	
	\item Partially collateralized: $P\{\tau>T\}$ = down-and-out(B) Binary Call(K) / df(T)
	
	Now the barrier level is set to represent the degree of collateralization.  A requirement for an Independent Amount would mean B$>$K, and a non-zero Threshold would imply B$<$K.
	
	\item Remargining frequency: $P\{\tau>T\}$ = discretely observed down-and-out(B) Binary Call(K) / df(T)
	
	We can add in observation frequency via the barrier option correction presented in \cite{Broadie1997a}.  
	
	\item Composite: we convert both discrete collateralizations into continuous barriers and use the higher of the two. This can be directly generalized to any number of discrete collateralizations at different levels of collateralization.
	
\end{itemize}
The Merton model is a degenerate undiscounted down-and-out Binary Call where the barrier level is zero.  The barrier correction in \cite{Broadie1997a} moves the barrier away by a factor of $\exp(\beta\sigma\sqrt{\Delta t})$ where $\beta\approx 0.5826$ and $\Delta t$ is the interval between observations.  There is both theoretical and experimental justification for the factor.  Of course we do not change the level of the final barrier.  Hence we have a single form for the survival probability that covers all five models:
\be
P\{\tau>T;V,B,K,T,\Delta T\} = e^{r T}\times
\begin{cases}
C_d(V,T,K) - \left( \frac{V}{\widehat{B}} \right)^{2\alpha} C_d(\widehat{B}^2/V,K)
&\mbox{if } K > \widehat{B} \\
C_d(V,T,\widehat{B}) - \left( \frac{V}{\widehat{B}} \right)^{2\alpha} C_d(\widehat{B}^2/V,\widehat{B})
&\mbox{if } K \le \widehat{B}
\end{cases}
\ee
where:

$K$ strike of Binary option;

$T$ maturity;

$C_d(V,T,K)$ Binary call option, starting value of underlying $V$;

$\widehat{B}$ adjusted barrier, $\widehat{B} = B e^{-\beta\sigma\sqrt{\Delta T}}$, \cite{Broadie1997a};

$\beta=-\zeta(1/2)\sqrt{2\pi}\approx 0.5826$, where $\zeta()$ is the Riemann zeta function;

\noindent
and:

\be
\alpha = \h \left(1 - \frac{r - D}{\sigma^2/2}\right);
\ee
where:

$r$ riskless rate, aka drift of GBM process of $V$;

$D$ dividend rate of $V$;

$\sigma$ volatility of GBM process of $V$.

\noindent
For the Composite model:
\begin{itemize}
	\item $\widehat{B} = \max_{i=1,\ldots,n}\{B_i e^{-\beta\sigma\sqrt{(\Delta T)_i}}\}$ where there are $n$ barriers;
	\item $K=\widehat{B}$.
\end{itemize}

\vskip3mm
The biggest issue with structural models is calibration which depends on the unobservable volatility of the underlying assets of the firm.  A convenient ersatz is often the stock price volatility despite its inconsistent nature.  A further refinement is to adjust the stock price volatility by the delta of the relevant Black-Scholes call option according to the moneyness of the option \cite{Schonbucher2003a} Section 9.4.3.  Here we use the simpler version.

Calibrating structural models to firm balance sheets in terms of their assets and liabilities can also be problematic \cite{Schonbucher2003a}.  We take two approaches here: we consider affect on longterm historical default probabities taken from \cite{Vazza2012a} (their Table 13); and alternatively affects on market-implied default probabilities using CDS spreads. 

\begin{table}
	\centering
\begin{tabular}{ccccc} \\
Rating & CDS 5Y & rec & S\&P 5Y default & S\&P as CDS\\ 
 & spread (bps) & \% & probability (bps) & spread (bps)  \\ 
\hline
 \text{A} & 90. & 38. & 66.  & 8.\\
 \text{BBB} & 130. & 38. & 230. & 29.\\
 \text{BB} & 290. & 37. & 861.  & 113.\\
 \text{B} & 510. & 36. & 2083.  & 299. \\
\end{tabular}
	\caption{Generic 5Y CDS spreads (as of Oct 2012) and longterm S\&P 5Y global cumulative default probabilities.  The final column are the S\&P default probabilities expressed as CDS spreads using the same recovery rates as the market CDS spreads.}
	\label{t:param}
\end{table}

\section{Results}

We first give the general setup and then the specific case studies, a non-bank case (range of Ratings and starting with zero collateralization) and a bank case (starting with a range of collateralization --- moderate to high --- considering only single-A Rating).

\subsection{Setup}

For simplicity we describe results in terms of default probabilities and in terms of CDS spreads.  CDS spreads are the common language of trading and often used within risk management.  We consider 5Y survival because the 5Y CDS is usually the most liquid point on the CDS curve.  Table \ref{t:param} gives the parameters.  

\begin{figure}[htbp]
	\centering
		\includegraphics[width=0.7\textwidth]{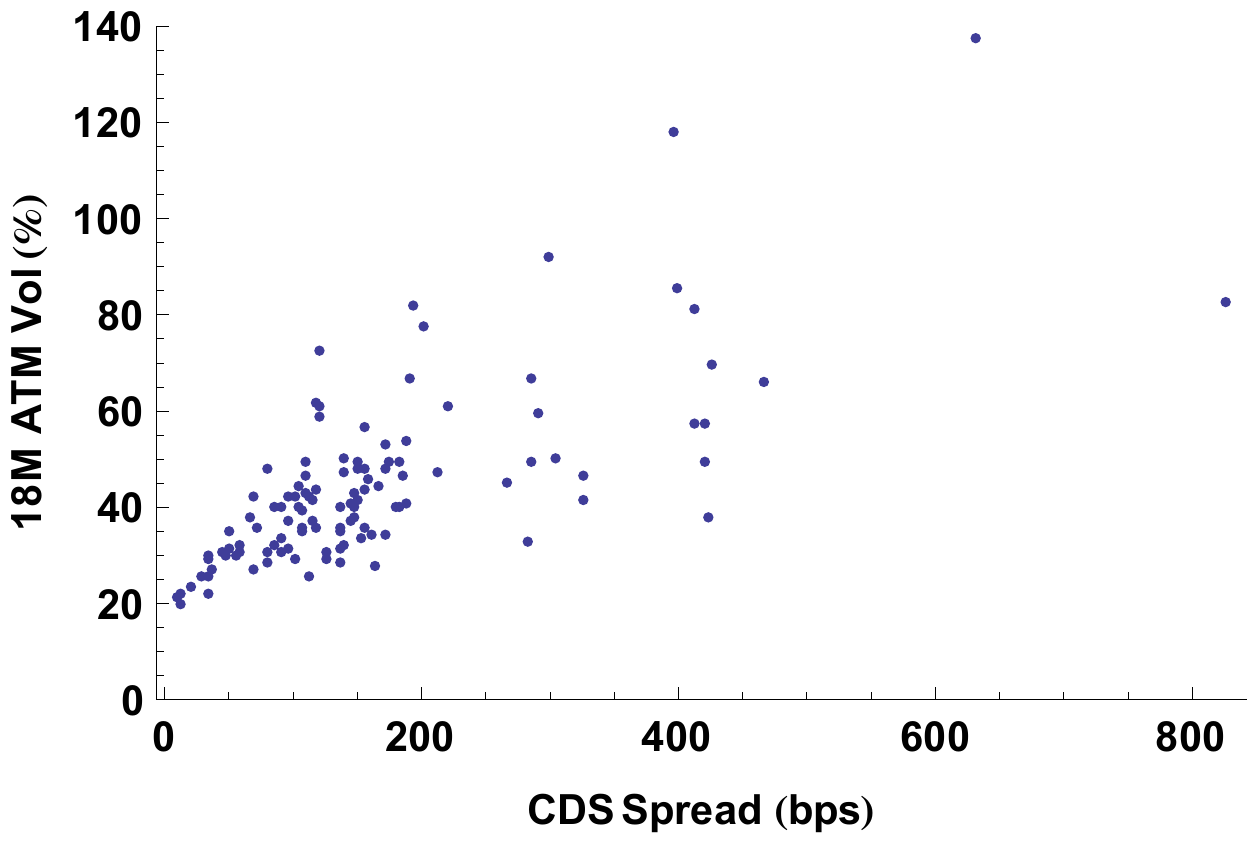}
	\caption{18M ATM volatilities of six US banks across the financial crisis (mid 2007 to end 2011, quarterly) versus their CDS spreads.  All currently rated A+/./- by S\&P (local long-term).}
	\label{f:CDSvol}
\end{figure}

The final parameter we need to choose is stock price volatility.  This is highly variable depending on whether there is a crisis or not.  Figure \ref{f:CDSvol} shows the average 18M market implied volatility versus CDS spread for six US banks all currently rated (local, long-term) A+/./- by S\&P (data from Bloomberg, tickers BAC, WFC, JPM, C, MS, GS).  18M market impled volatility is often the longest maturity that is liquid.  At longer maturities there is often little smile.  Both CDS and implied volatility can change significantly during a crisis before rating agencies have time to react.  Based on Figure \ref{f:CDSvol} we analyse a range of stock volatilities from 20\%\ to 80\%\ for all the different CDS spreads in Table \ref{t:param}.

\subsection{Case Studies}

We pick two situations to analyse, a typical non-bank trading with banks (e.g. an insurance company trading with a major bank) and a bank setup.  The key point is that non-banks will usually trade uncollateralized because they are cash-flow optimizers, on the other hand banks will typically already have a large to very large proportion of their trades collateralized.  

In both cases we analyse the result of increasing collateralization relative to their starting positions.  Of course, if the non-bank trading with a bank defaults then that can cause problems for the bank is the exposure is high enough.  This was the potential situation with AIG when it hit its rating trigger requiring it to post collateral.  The banks on the other side of its trades might have had serious problems if AIG had not been bailed out.  Conversely, if there had been no triggers then it is possible that AIG might not have required bailing out to the same extent, and the financial crisis did end.  The recently reported case of a large European bank and Canadian insurance companies might be an example \cite{Lokey2012a}.  Hindsight is 20:20 and we have only one historical record so these speculations should be taken as simple examples of possibilities.

\subsubsection{Non-Bank}

We consider four rating setups (B to A) from both marked-implied and historical default probabilities.  Marked-implied default probabilities are as of October 2012 and historical use 1981-2011.

\begin{figure}[htbp]
	\centering
		\includegraphics[width=0.95\textwidth]{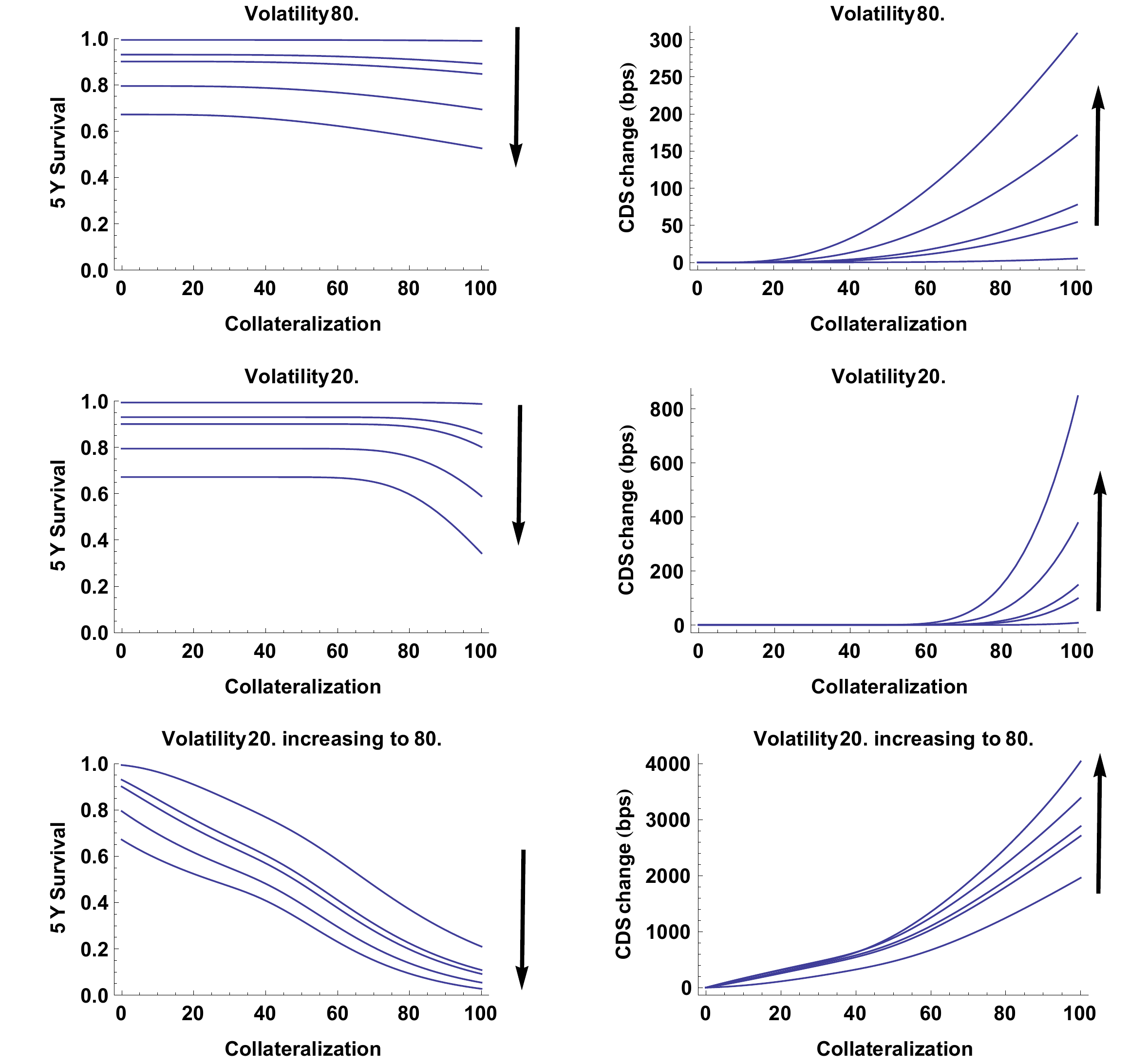}
	\caption{Non-Bank example, collateralization is increased from zero to 100\%.  Effect on 5Y survival is on the LHS and increase in CDS spreads on the RHS.  Each line is a different starting CDS spread \{8, 90, 130, 290, 510\}bps (lowest CDS spread has highest survival probability), parameters are from Table \ref{t:param}. {\bf arrows} show direction of increasing starting CDS spread for the different lines.  Rows show different equity volatility setups.  In the last row equity volatility is increased linearly with increasing collateralization.  This mimics sliding triggers in a CSA as equity volatility increases as a proxy for downgrades.  Curves are in opposite order on the LHS to the RHS, i.e. the biggest CDS changes arise from the lowest survival probabilities.}
	\label{f:non-bank}
\end{figure}

Figure \ref{f:non-bank} shows the effect as collateralization is increased from zero to 100\%\ for starting CDS spreads of \{8, 90, 130, 290, 510\}bps so as to cover the range of  both the market-implied and the S\&P long-term cases.  If there is already a high-volatility situation (80\%) then the effect of increaseing collateralization, even from zero to 100\%\ is relatively mild, with a maximum CDS spread increase of 300bps.  If there is a low volatility situation (20\%) then increasing collateralization has a larger effect because the terminal default barrier is higher to start with. The maximum effect is now 800bps. 

When we look at collateralization being increased at the same time as volatility increasing we see a qualitatively different picture.  There is an immediate increase in equivalent CDS spreads and it is hugely more significant with a maximum incrase of 4000bps (40\%).  This is the typical case for triggers.  As the market situation, and the situation of the non-bank worsens more collateral is required.  From this perspective the feedback is clear.  Whilst this has been qualitatively clear we now show quantitatively just how bad triggers can be.  Effectively they defeat their purpose.

\subsubsection{Bank}

We consider one rating (single A), using a CDS spread of 90bps (market implied), and a range of starting collateralizations: 50\%\, 70\%\ and 90\%.  Actual proportions are difficult to calculate without internal data so we simply look at a range.  One could imagine universal banks lower in the range and investment banks higher in the range. 

\begin{figure}[htbp]
	\centering
		\includegraphics[width=0.95\textwidth]{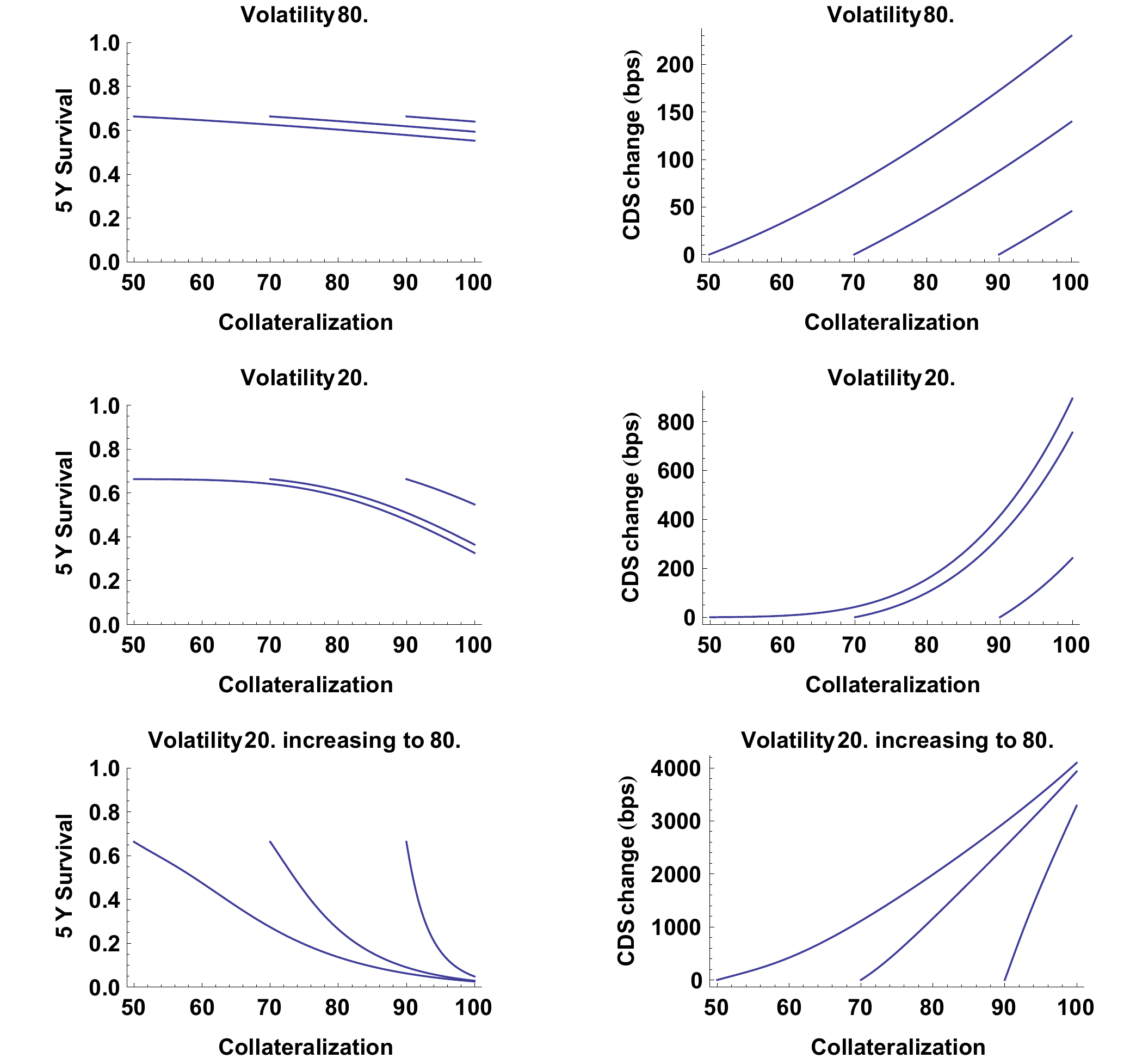}
	\caption{Bank example, collateralization is increased from 50\%\, 70\%\ and 90\%\ to 100\%.  Effect on 5Y survival is on the LHS and increase in CDS spreads on the RHS.  Each line is a different starting collateralization which may be read by looking at their starting point (left-end-point) on the horizontal axis.  Rows show different equity volatility setups.  In the last row equity volatility is increased linearly with increasing collateralization.  This mimics sliding triggers in a CSA as equity volatility increases as a proxy for downgrades.  Curves are in opposite order on the LHS to the RHS, i.e. the biggest CDS changes arise from the lowest survival probabilities.}
	\label{f:bank}
\end{figure}

Qualitatively Figure \ref{f:bank} is similar to Figure \ref{f:non-bank} which shows the non-Bank case.  Starting with the high volatility means that there is little effect of increasing collateralization although it is still significant relative to the starting CDS level of 90bps.  Starting with low volatility means that both barriers are higher (unlike the non-bank case the bak case starts with non-zero collateralization, i.e. a non-zero Black-Cox default barrier) so increases in collateralization have a bigger effect.

As with the non-bank case, increasing collateral requirements at the same time as increases in volatility has a drastic effect.  The bank setup starts from a non-zero Black-Cox default barrier and so volatility increases have an equivalent, or greater, effect as compared with the non-bank case.

\section{Discussion}

Changes in collateralization have been implicated in significant default (or near-default) events during the financial crisis, most notably with AIG.  We have developed a framework for quantifying this effect based on moving between Merton-type and Black-Cox-type structural default models.  Our framework leads to a single equation that emcompasses the range of possibilities, including collateralization remargining frequency (i.e. discrete observations).

Collateralization reveals asset value, thus mark-to-market volatility can lead to default even in cases where the firm would recover later.  It is at once a philosophical and immediately relevant question as to whether such "virtual" defaults are important or not.  Forcing collateralization as political and regulatory policy makes these virtual defaults actual.

In our results we examined the effect of changes in collateralization for a typical non-bank counterparty starting from zero collateralization and for a typical bank starting from various degrees of collateralization.  We showed that there are significant increases in default probability, as measured by 5Y CDS spreads, in both low-volatility and high-volatility situations.  However, increasing collateralization enhances default probability to an extreme degree when moving from a low volatility situation to a high volatility situation.  Thus. at least qualitatively. we reproduce the AIG event.

Technically our model can be enhanced in several directions.  One angle would be to use structural default models  which also have good short-term behaviour (several are mentioned in \cite{Schonbucher2003a}) and so examine the effect of increasing collateralization on CDS term structure.  Another angle would be to attempt a calibration to individual firms and their specific degree of collateralization.  Including volatility of liabilities as well as assets provides another avenue for future research.

We have not considered possible deterioration of firm assets as a function of their volatility or of firm CDS spread.  Thus we have not captured any potentially systematic component of default risk correlated with idiosyncratic risk.  This is explored in Chapter 9 of \cite{Kenyon2012a} where simultaneous changes in both asset and firm CDS spread are explored  (obviously some regions are infeasible).  In this respect our modeling of the enhancement of default by collateralization is conservative.  Although triggers appear very bad in terms of increasing default risk, we have underestimated their likely effect during a crisis.  We leave exploration of systematic and idiosyncratic risk, within our framework, for future research.

There is also a proposal to make standard initial margins (IM) compulsory for bilateral trades.  IMs are required by central counterparties (CCPs) but the bilateral IM proposal is higher \cite{BCBS-226,BCBS-227}.  This has led to concern about the quantity of collateral required \cite{ISDA-2012-Margin}.  In our framework IM is just a move of the default barrier because IM makes some assets unavailable for collateral posting as they are already committed.

This paper shows that increases in collateralization, by exposing entities to daily mark-to-market volatility of their assets, enhance default probability.  This quantifies the well-known problem with collateral triggers.  Furthermore our models can be used to quantify the degree to which central counterparties, whilst removing one credit risk transmission channel, increase default risk.  

\bibliographystyle{chicago}
\bibliography{kenyon_general}
\end{document}